\documentclass[12pt]{article}

\usepackage{amsfonts,amsmath}
\usepackage{latexsym}

\usepackage{epsfig}
\usepackage{psfrag}

\pdfoutput=1

\usepackage{epsfig}
\newcommand{\rf}[1]{(\ref{#1})}
\newcommand{\oh}{\frac{1}{2}}

\newcommand{\bea}{\begin{eqnarray}}
\newcommand{\eea}{\end{eqnarray}}
\newcommand{\beas}{\begin{eqnarray*}}
\newcommand{\eeas}{\end{eqnarray*}}
\newcommand{\beqs}{\begin{displaymath}}
\newcommand{\eeqs}{\end{displaymath}}

\newcommand{\ben}{\begin{equation}}
\newcommand{\een}{\end{equation}}

\newcommand{\bdm}{\begin{displaymath}}
\newcommand{\edm}{\end{displaymath}}

\newcommand{\pa}{\partial}

\def\void{}
\def\labelmark{\marginpar{\small\labelname}}

\newenvironment{formula}[1]{\def\labelname{#1}
\ifx\void\labelname\def\junk{\begin{displaymath}}
\else\def\junk{\begin{equation}\label{\labelname}}\fi\junk}%
{\ifx\void\labelname\def\junk{\end{displaymath}}
\else\def\junk{\end{equation}}\fi\junk\labelmark\def\labelname{}}

{\ifx\void\labelname\def\junk{\end{array}\end{displaymath}}
\else\def\junk{\end{array}\right.\end{equation}}
\fi\junk\labelmark\def\labelname{}\def\junk{}
}

\newcommand{\beq}{\begin{formula}}
\newcommand{\eeq}{\end{formula}}
\newcommand{\beqv}{\begin{formula}{}}

\begin{document}

\hfill

\addtolength{\baselineskip}{0.5\baselineskip}
 \topmargin 0pt
 \oddsidemargin 5mm
 \headheight 0pt
 \topskip 0mm


\begin{center}

{\Large \bf Exponential bounds on the  \\
number of causal triangulations}

\medskip
\vspace{1 truecm} 


\vspace{1 truecm}

{\bf Bergfinnur Durhuus}\footnote{email: durhuus@math.ku.dk}

\vspace{0.4 truecm}

Department of Mathematical Sciences

University of Copenhagen, Universitetsparken 5

DK-2100 Copenhagen \O, Denmark

 \vspace{1.3 truecm}

{\bf Thordur Jonsson}\footnote{email: thjons@hi.is}

\vspace{0.4 truecm}

Division of Mathematics, The Science Institute

University of Iceland, Dunhaga 3

IS-107 Reykjavik, Iceland

\end{center}
 \pagestyle{empty}

\hfill

\noindent {\bf Abstract.} We prove that the number of combinatorially distinct causal 3-dimensional triangulations homeomorphic to 
the 3-dimensional sphere is bounded by an exponential function of the number of tetrahedra. It is also proven that the number of 
combinatorially distinct causal 4-dimensional triangulations homeomorphic to the 4-sphere is bounded by an exponential function 
of the number of 4-simplices provided the number of all combinatorially distinct triangulations of the 3-sphere is bounded by an 
exponential function of the number of tetrahedra.

 \newpage
 \pagestyle{plain}

\section{Introduction}  Random triangulations have been used for over 30 years to construct discrete models of 2-dimensional 
quantum gravity.  For a review of early work in this field, see \cite{book}.  In \cite{adj1} the generalization of these 
models to three dimensions was considered and it was pointed out that one needs bounds on the number of combinatorially distinct 
triangulations of a given fixed topology in order for the models to be well defined.  More precisely: If $N(V)$ is the number of 
distinct triangulations of the 3-sphere, say, by $V$ tetrahedra we need a bound of the form
\begin{equation}\label{b}
N(V)\leq C^V\,,
\end{equation}
for some constant $C$. It is easy to deduce from (\ref{b}) the corresponding bound for triangulations of the 3-ball or, 
more generally, the 3-sphere with a number of 3-balls removed.  

In \cite{dj} the notion of locally constructible triangulations was introduced and such triangulations 
of $S^3$ were shown to obey the 
bound \rf{b}.  Subsequently it was proven that not all triangulations of 
$S^3$ are locally constructible \cite{bz} and a proof of \rf{b} is still missing.   See \cite{ce} for a recent 
result which gives a new sufficient condition for \rf{b} to hold.   Monte Carlo simulations 
of the 3-dimensional gravity models indicate that a bound of the form \rf{b} is valid, see \cite{mc1,mc2}.   

An alternative approach to discretized quantum gravity in two and higher dimensions is the use of so-called causal triangulations.   
This is the class of triangulations that we are concerned with in the bulk of this paper. A general definition is given in 
Section~\ref{prelim}. In the case of 3 dimensions they can be described as follows: We first introduce the notion of 
a {\it causal slice} which is a triangulation of $S^2\times [0,1]$ with the property that all the vertices lie on the boundary and 
every tetrahedron has a least one vertex in each of the two boundary components.  If $K$ is a causal slice then its boundary $\pa K$
consists of two triangulations of the 2-sphere.  
We choose to label the boundary components and call one of them the in-boundary and the other one the out-boundary.   
A {\it causal triangulation} $K$ is then defined by a sequence $K_1, K_2,\ldots ,K_N$ of causal slices where $K_j$ has 
in-boundary $\Sigma^j_{\rm in}$ and out-boundary $\Sigma^j_{\rm out}$
such that $K^j$ and $K^{j+1}$ intersect in $\Sigma^j_{\rm out}  = \Sigma^{j+1}_{\rm in}$ for $j=1,2\ldots ,N-1$, and $K^i$ 
and $K^j$ are disjoint otherwise.
Then  $K$ is a triangulation of $S^2\times [0,1]$ with boundary $\pa K= \Sigma^1_{\rm in}\cup   
\Sigma^N_{\rm out} \equiv \Sigma_{\rm in}\cup\Sigma_{\rm out}$.   

If $T$ is a triangulation of the 3-dimensional ball with one interior vertex and boundary $\pa T= \Sigma_{\rm in}$ then we obtain 
another triangulation of the ball by attaching $T$ to the causal triangulation $K$ along the in-boundary.  Such triangulations are often referred to as 
causal triangulations of the 3-ball.  Obviously we can close up at the other end in the same way
 and obtain a triangulation of $S^3$.

One can think of the triangulated 
2-spheres in the boundaries of the causal slices as the "space" and the graph distance from 
$\Sigma_{\rm in}$ as a discrete "time" coordinate so space-time is in this picture 
foliated by a sequence of 2-spheres which are connected by causal slices. This approach was introduced in \cite{al1} for 2-dimensions 
and generalized to 3-dimensions in \cite{al2}.  For a while it was hoped that this model could be solved 
exactly in 3 dimensions because it can be 
mapped onto a certain matrix model \cite{al3} but this turned out to be overly optimistic.
The 4-dimensional models are still under active investigation \cite{al4} but so far almost all results are numerical.  

For the causal triangulations 
to make sense in 3 dimensions as a model for quantum gravity an exponential bound of the 
form $\rf{b}$ is also needed.  The main purpose of this paper is to provide such a bound.  
Our method is partly inspired 
by the matrix model approach \cite{al3} which codes the information about the causal slices 
in certain graphs whose structure will be described in the next section. 

We introduce the notion of {\it generalized causal slices} 
which are defined in the same way 
as causal slices except they are not required to be homeomorphic  
to $S^2\times [0,1]$.  Instead they are required to be simplicial manifolds with two boundary components and this 
definition extends in a straightforward way to higher dimensions.   Generalized causal triangulations are then 
defined as a sequence of generalized causal slices with boundaries identified as described above.

The principal results of this paper are the following:
\begin{itemize}
\item[{\bf A}] The number $N_3(V)$ of combinatorially distinct 3-dimensional causal 
triangulations made up of $V$ tetrahedra satisfies the
inequality \rf{b}.    

\item[{\bf B}] The two boundary components of generalized $3$-dimensional 
causal triangulations are necessarily homeomorphic to each other 
and there is an exponential bound like \rf{b} on the number of distinct generalized causal triangulations with boundaries of a fixed genus. 

\item[{\bf C}] Assuming the bound \rf{b} for arbitrary triangulations of $S^3$, it is proven 
that the number of distinct causal triangulations of $S^3\times [0,1]$ is bounded by an exponential function of the number of 4-simplicies.

\end{itemize}

In the next section we establish 
some preliminary results on causal slices.  In Section 3 we prove {\bf A}, discuss 
generalized causal triangulations and establish {\bf B}.   It follows from {\bf A}, as we show in Section 3,  that the number 
$N_3(V, \Sigma_{\rm in},\Sigma_{\rm out})$ of causal triangulations of $S^2\times [ 0,1]$ made up of $V$ tetrahedra
with fixed boundaries $\Sigma_{\rm in}$ and $\Sigma_{\rm out}$
is given by
\begin{equation}\label{ex}
N_3(V,\Sigma_{\rm in},\Sigma_{\rm out})=e^{\beta V+o(V)},
\end{equation}
where $\beta$ is independent of $\Sigma_{\rm in}$ and $\Sigma_{\rm out}$.
In Section 4 we extend some of the results from Section 3
to the 4-dimensional case and prove {\bf C}.

\section{Preliminaries}\label{prelim}
In this section we establish some notation and give a general definition of the principal objects under study in this paper,  causal slices 
and triangulations, as well as some associated objects.

Recall that an abstract simplicial complex $K$ is defined by its vertex set $K^0$, which here is assumed to be finite, and a set 
of subsets of $K^0$, called simplices, such that if $\sigma$ is a simplex in $K$ and $\sigma'\subset \sigma$ then $\sigma'$ is also a 
simplex in $K$ (see \cite{plt}). If $\sigma$ is a simplex in $K$ containing $p+1$ vertices it is called a $p$-simplex and the set of $p$-simplices 
will be denoted by $K^p$.  We denote by $|K^p|$ the number of $p$-simplicies in $K$.  
If every simplex in $K$ is contained in some $D$-simplex we say that
 $D$ is the dimension of $K$.
Given two abstract simplicial complexes $K$ and $K_1$, a bijective map $\psi:K_1^0\to K^0$ is called a 
\emph{combinatorial isomorphism} if it induces a bijection of simplices. Obviously, combinatorial isomorphism is an equivalence relation.

A (geometric) \emph{realization} of an abstract simplicial complex $K$ is a map 
$$
\phi:K^0\to \mathbb R^n
$$ 
for some $n$ such that $\phi(\sigma)$ is 
an affinely independent set for every simplex $\sigma$ in $K$ and 
$$
\mbox{conv}\,\phi(\sigma)\cap \mbox{conv}\,\phi(\sigma') = \mbox{conv}\,\phi(\sigma\cap\sigma')
$$
for all simplices $\sigma, \sigma'$ in $K$.   Here, $\mbox{conv}\,\phi (\sigma )$ 
is the convex hull of the set $\phi (\sigma )$.
Hence, the convex spans of the images of abstract simplices in $K$ define a simplicial complex 
in $\mathbb R^n$ that we shall denote $K_\phi$. It is well known (see, e.g., \cite{seifert}) that any abstract simplicial complex has a geometric 
realization for $n$ sufficiently large and that the homeomorphism class of 
$$
K'_\phi = \bigcup_{\sigma\in K} \mbox{conv}\,\phi(\sigma)
$$
does not depend on the realization $\phi$. If $K'_\phi$ is a manifold we call $K$ a manifold.  In the following 
we assume that all manifolds are connected. 
Note also that if $\psi$ is a combinatorial 
isomorphism as defined above and $\phi$ is a realization of 
$K$ then $\phi\circ \psi$ is a realization of $K_1$ and we may speak about $\phi$ 
as a realization of the combinatorial isomorphism class of $K$. Frequently, we shall not distinguish between $K$, its combinatorial 
isomorphism class, and its geometric realizations. Similarly, if $K$ is a $D$-dimensional manifold, the boundary complex $\partial K$ 
is defined in the standard manner as an abstract simplicial complex consisting of all the 
$(D-1)$-simplices (together with their 
subsimplices) contained in only one $D$-simplex. A realization of $K$ gives in an obvious way rise to a realization 
of $\partial K$ in the same Euclidean space.

 From now on we shall assume that $K$ is a manifold of dimension $D$.   For the purpose of defining the causal slices
 we further assume that $K^0$ is  divided into two classes whose 
vertices we will call red  and blue.  These will be the vertices of the 
in- and out-boundary components discussed in the Introduction.  
By definition, a $p$-simplex whose vertices all have the same 
colour inherits the colour of the vertices.  We will require that no $D$-simplex has all its vertices of the same colour.  
The $D$-simplices therefore fall into $D$ classes determined by the number $k$ of red vertices, and hence $D+1-k$ blue vertices, which 
we say are of type $(k,D+1-k)$, see Fig.\ 1. 

\begin{figure}[t]
  \begin{center}
   \includegraphics[width=12truecm]{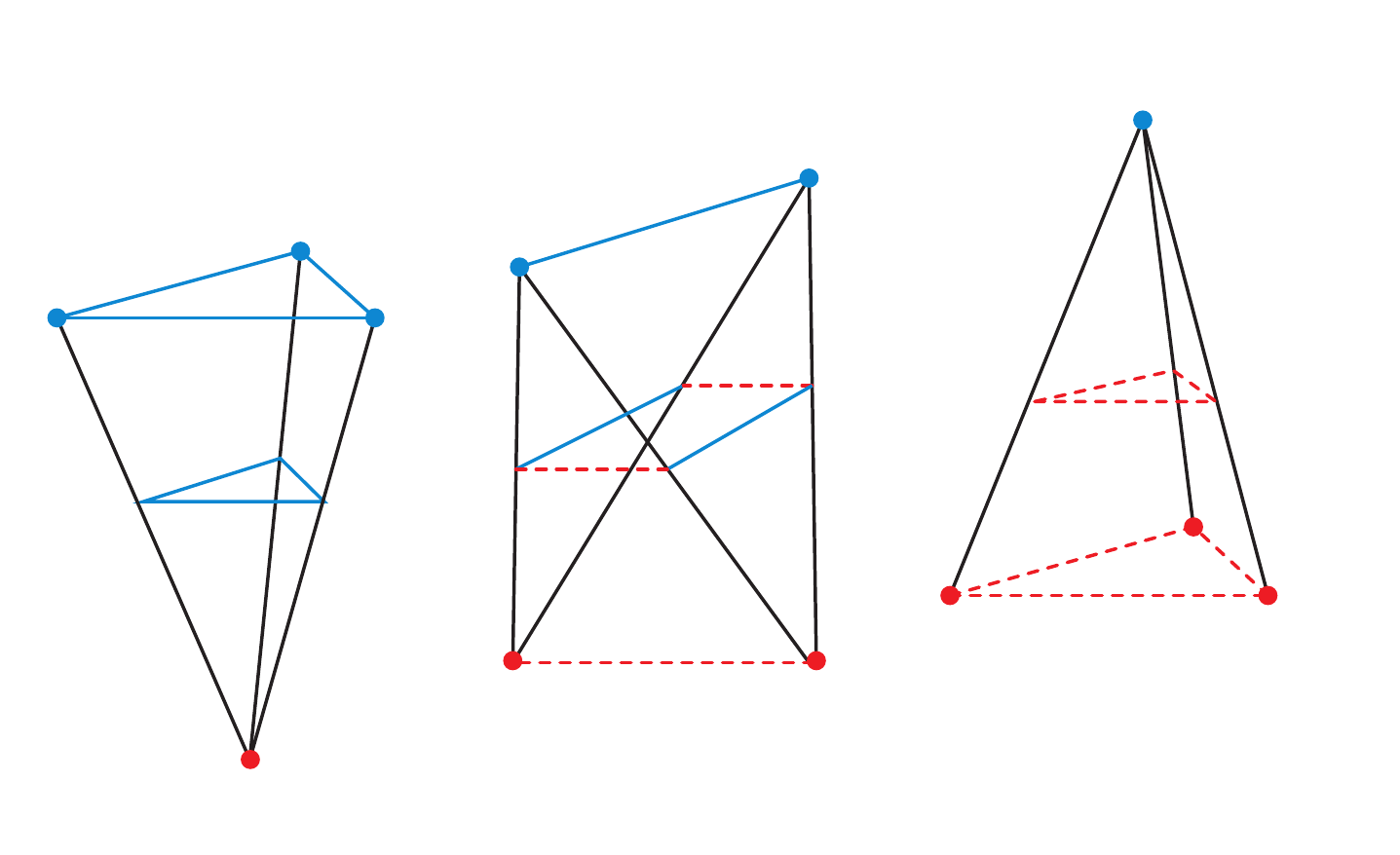}
    \caption{The three different types of $D$-simplicies that arise for $D=3$.  The full lines are blue and 
    the broken ones red.}
    \label{fig1}
      \end{center}
      \end{figure}

\medskip
\noindent
{\bf Definition 1.}  A {\it D-dimensional causal slice} $K$ is an abstract simplicial complex as described above which satisfies the following 
conditions:
\begin{enumerate}

\item[(i)] $K$ is homeomorphic to the cylinder $S^{D-1}\times [0,1]$.  

\item[(ii)] All mono-coloured simplices of $K$ belong to the boundary $\pa K$, such that the red ones belong to one boundary 
component $\partial K_{red}$ and the blue ones to the other component $\partial K_{blue}$.  
 \end{enumerate}
A {\it causal triangulation} of dimension $D$ is an abstract simplicial complex of the form 
$$
M = \bigcup_{i=1}^N K^i\,,
$$
where each $K^i$ is a causal slice of dimension $D $ such that  $K^i$ is disjoint from $K^j$ for 
$i\neq j$ except that $\partial K^{i}_{blue} = \partial K^{i+1}_{red},\, i= 1,\dots, N-1,$ as uncoloured 
abstract simplicial complexes. The boundary components of $\partial K^1_{red}$ and $\partial K^N_{blue}$ of $M$ will be denoted 
$\Sigma_{in}$ and $\Sigma_{out}$, respectively.  

\bigskip

With the colouring conventions described above, we note that (ii) in Definition 1
 is equivalent to the requirement in the Introduction (where we had $D=3$) that all 
vertices belong to the boundary and no tetrahedron has all its vertices in the same boundary component. 
There is an obvious notion of combinatorial isomorphism of causal slices respecting the colouring, and 
we shall denote by ${\cal C}{\cal S}_D$ the set of combinatorial equivalence classes of $D$-dimensional causal slices. Similarly, we denote 
by ${\cal C}_D$ the set of combinatorial equivalence classes of causal triangulations of dimension $D$. 

Let $K$ be a causal slice of dimension $D$ and consider a realization $\phi$ of $K$ in $\mathbb R^n$.
Denote the boundary components of $K'_\phi$ corresponding to $\partial K_{red}$ and $\partial K_{blue}$ by $\partial K_{red}'$ and 
$\partial K_{blue}'$, respectively.
Given $x\in K'_\phi$ there is a $D$-simplex $s$ in $K_\phi$ containing $x$.  If we let $\{r_i\}$ and $\{b_j\}$ 
denote the red and blue vertices, respectively, in $s$, we can express $x$ in a unique 
way as a convex combination
\begin{equation}\label{convex}
x=\sum_i\mu_ir_i+\sum_j\lambda_j b_j,
\end{equation}
where $\mu_i,\lambda_j\geq 0$ and $\sum_i\mu_i+\sum_j\lambda_j =1$.  Depending on the colour type of $s$,
  $i$ runs from $1$ to $k$ and $j$ runs from $1$ to $D+1-k$, $1\leq k\leq D$. 
 We define the \emph{height function} 
 $h:K'_\phi\to [0,1]$ by
\begin{equation}\label{hdef}
h(x)=\sum_j\lambda_j.
\end{equation}
It is easy to check that this function is well defined, i.e., $h(x)$ does not depend on the choice of $s$ containing $x$, and it 
is a simplicial map, see \cite{plt}. Evidently, $h(x)=0$ if and only if 
$x\in \partial K_{red}'$ and $h(x)=1$ if and only if $x\in \partial K_{blue}'$.  Moreover, 
if $0<h(x)<1$, then there is a unique line segment $[x_r,x_b]$ with endpoints $x_r\in\partial K_{red}'$ and $x_b\in\partial K_{blue}'$ that contains $x$ 
and is contained in some $D$-simplex in $K'_\phi$.  With the notation introduced above we have
\begin{equation}\label{hf2}
x_r={1\over 1-h(x)}\sum_i\mu_ir_i,~~~x_b={1\over h(x)}\sum_j\lambda_jb_j.
\end{equation}
We now consider the section in $K'_\phi$ consisting of points at height $\oh$:
\begin{equation}\label{section}
S_K=\{ x\in K'_\phi: h(x)=\oh\}.
\end{equation}
We will refer to $S_K$ as the \emph{midsection} of $K'_\phi$. Actually, we shall consider $S_K$ as a coloured cell complex in 
the following sense, that we now describe in detail only for the cases $D=3$ and $D=4$. We refer to \cite{plt} for a definition of cell complexes.

Consider first the case $D=3$.  
If $s$ is a tetrahedron in $K_\phi$ of type $(k,4-k)$, $k\in\{ 1,2,3\}$, then $s\cap S_K$ is a cell with
$k(4-k)$ corners.   If $k=1$ or 3, then the cell is a triangle whose edges we declare to be red if $k=3$ but 
blue if $k=1$.  If $k=2$ then the cell $s\cap S_K$ is a quadrangle, i.e., a product of two 1-simplicies. The edges of the quadrangles 
are coloured by the same colour as the colour of the boundary edge in $\partial K_{red}'$ or $\partial K_{blue}'$ of the triangle 
to which they belong.  
In particular, opposite sides of the quadrangles have the same colour, see Fig.\ 2. 

\begin{figure}[h]
  \begin{center}
     \includegraphics[width=8truecm]{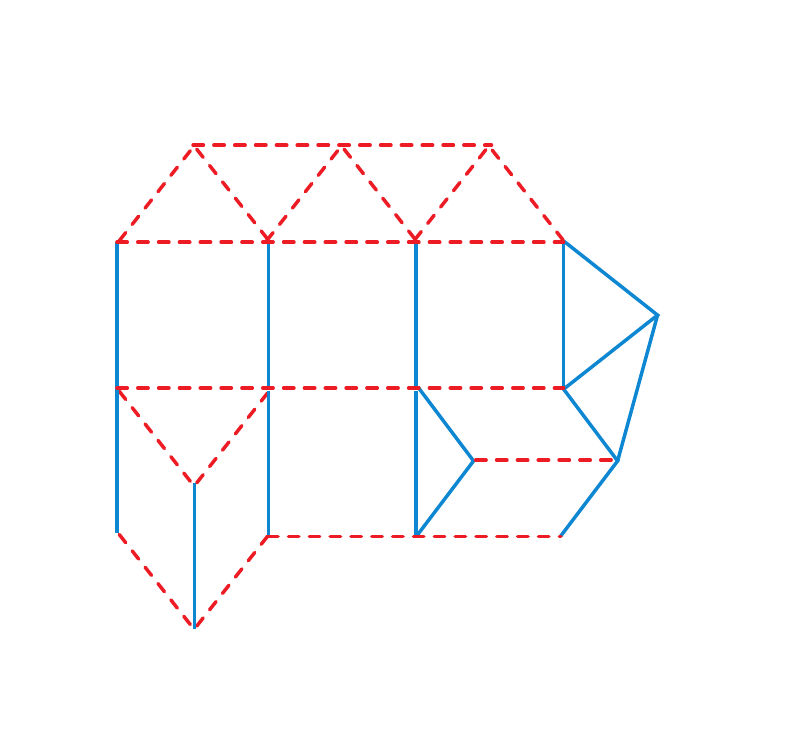}
	        \caption{A part of a midsection of a 3-dimensional causal slice.}
    \label{fig2}
      \end{center}
      \end{figure}

This defines  $S_K$ as a 2-dimensional cell 
complex with coloured edges. We shall also use $S_K$ 
to denote its combinatorial isomorphism class, where a combinatorial 
(or abstract in the terminology of \cite{plt}) isomorphism $\varphi:S_{K_1}\to S_{K_2}$ is defined as 
a bijective map from the 0-cells of $S_{K_1}$ onto those of $S_{K_2}$ such that $\{v_1,\dots,v_k\}$ is the set of corners of a 
cell in $S_{K_1}$ if and only if $\{\varphi(v_1),\dots,\varphi(v_k)\}$ is the set of corners of a cell in $S_{K_2}$ 
for $k=2,3,4$ and, in addition, $\varphi$ preserves the colouring of edges.  Note that changing the hight of the midsection to 
any $h\in (0,1)$ does not affect its combinatorial equivalence class but at $h=0$ and $h=1$ the section collapses to the red and 
blue boundaries, respectively.

In case $D=4$ there are four types of coloured cells. If $s$ is a $4$-simplex of type $(4,1)$ or $(1,4)$ then $s \cap S_K$ is a 
tetrahedron coloured red or blue, respectively. On the other hand, if $s$ is of type $(3,2)$ or $(2,3)$ then $s\cap S_K$ is a 
prism $\Delta\times [0,1]$, 
where $\Delta$ is a triangle. In the former case, the edges of the two triangles in the boundary 
are coloured red while the three 
remaining edges are coloured blue, and vice versa in the latter case (see Fig.\ 3). This defines $S_K$ as a 3-dimensional cell complex 
with coloured edges.  

\begin{figure}[h]
  \begin{center}
  \includegraphics[width=11truecm]{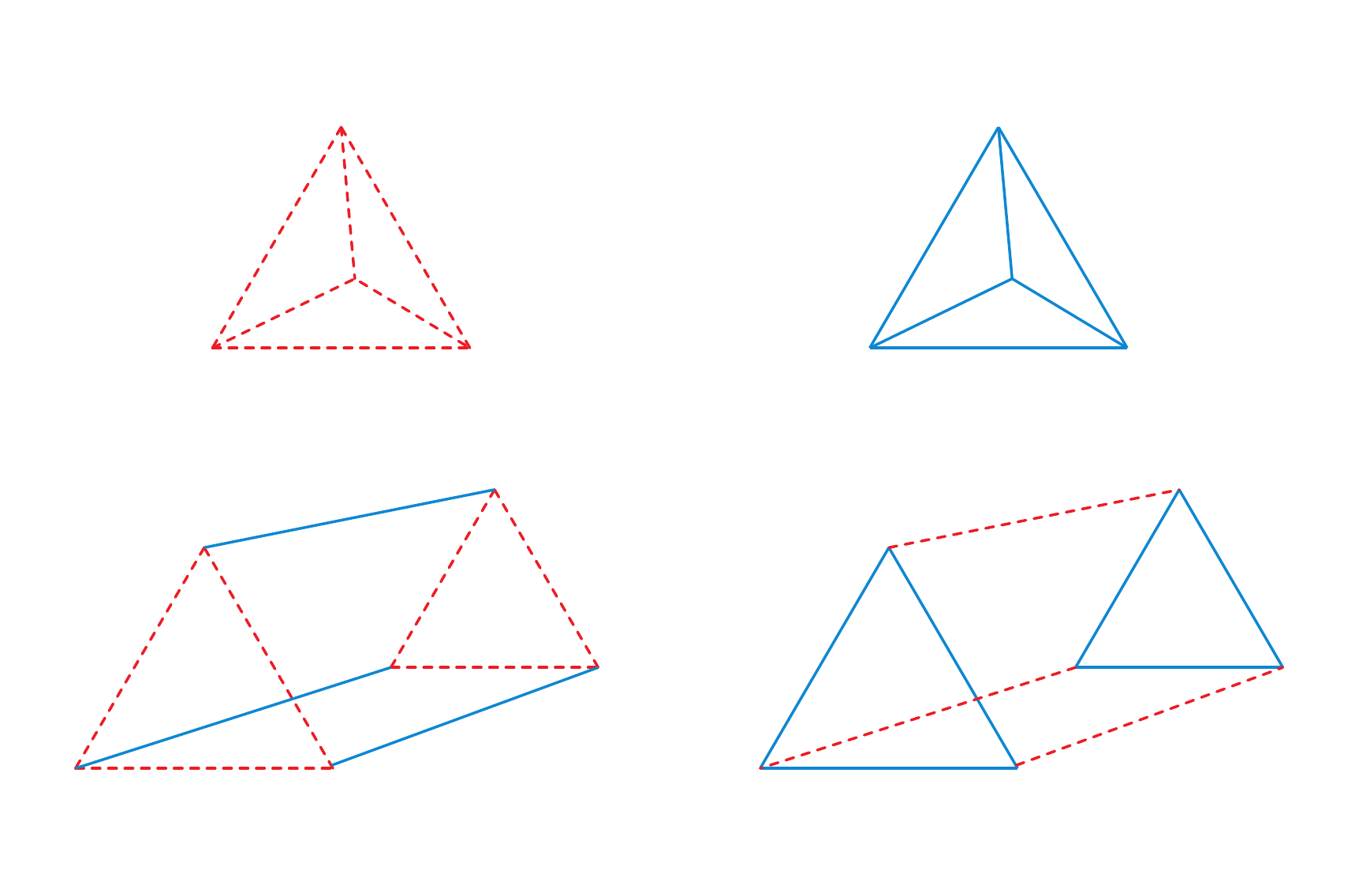}
  \caption{The four different coloured cells that occur in the midsection in the 4-dimensional case.}
    \label{fig3}
      \end{center}
      \end{figure}

 While we need a realization of $K$ in 
order to define the midsection, it is essentially independent of which realization is chosen
as decribed in the following Lemma.  This justifies the notation $S_K$ for the midsection.

\medskip
\noindent
{\bf Lemma 1.}   {\it The midsection $S_K$ of a causal slice $K$  is a closed $(D-1)$-dimensional manifold. Combinatorially 
isomorphic causal slices give rise to combinatorially isomorphic midsections.}

\medskip
\noindent
{\bf Proof.} 
Suppose we have a given realization $K_\phi$ of $K$ with a midsection $S_K$.
We have that $\frac 12$ is a regular value of the height function $h$, from which it 
follows that $S_K$ is a closed $(D-1)$-dimensional manifold (see, e.g., \cite{will} Sections 1.3 and 4.2). 

If $\psi$ is a combinatorial isomorphism between abstract simplicial complexes $K_1$ and $K_2$ there is a canonical piecewise linear 
homeomorphism between any two realizations $K_{\phi_1}$ and $K_{\phi_2}$  of them, which obviously restricts 
to a piecewise linear homeomorphism between the corresponding midsections $S_{K_{\phi_1}}$ 
and $S_{K_{\phi_2}}$ which in turn induces a combinatorial isomorphism between $S_{K_{\phi_1}}$ and $S_{K_{\phi_2}}$.  
This proves the last statement of the Lemma. \hfill $\square$

\medskip

The preceding Lemma allows us to define a mapping $\pi$ on ${\cal C}{\cal S}_D$ by setting 
$$
\pi(K) = S_K\,,
$$
such that $\pi$ takes values in the set of $(D-1)$-dimensional closed coloured 
cell-complexes as described above for the 
cases $D=3$ and $D=4$. Straightforward generalizations can be given for higher dimensions 
but we shall not consider these cases 
further in this paper. It is rather easy to see that $\pi$ is not a surjective map, see Fig.\ 4. 

\begin{figure}[h]
  \begin{center}
     \includegraphics[width=10truecm]{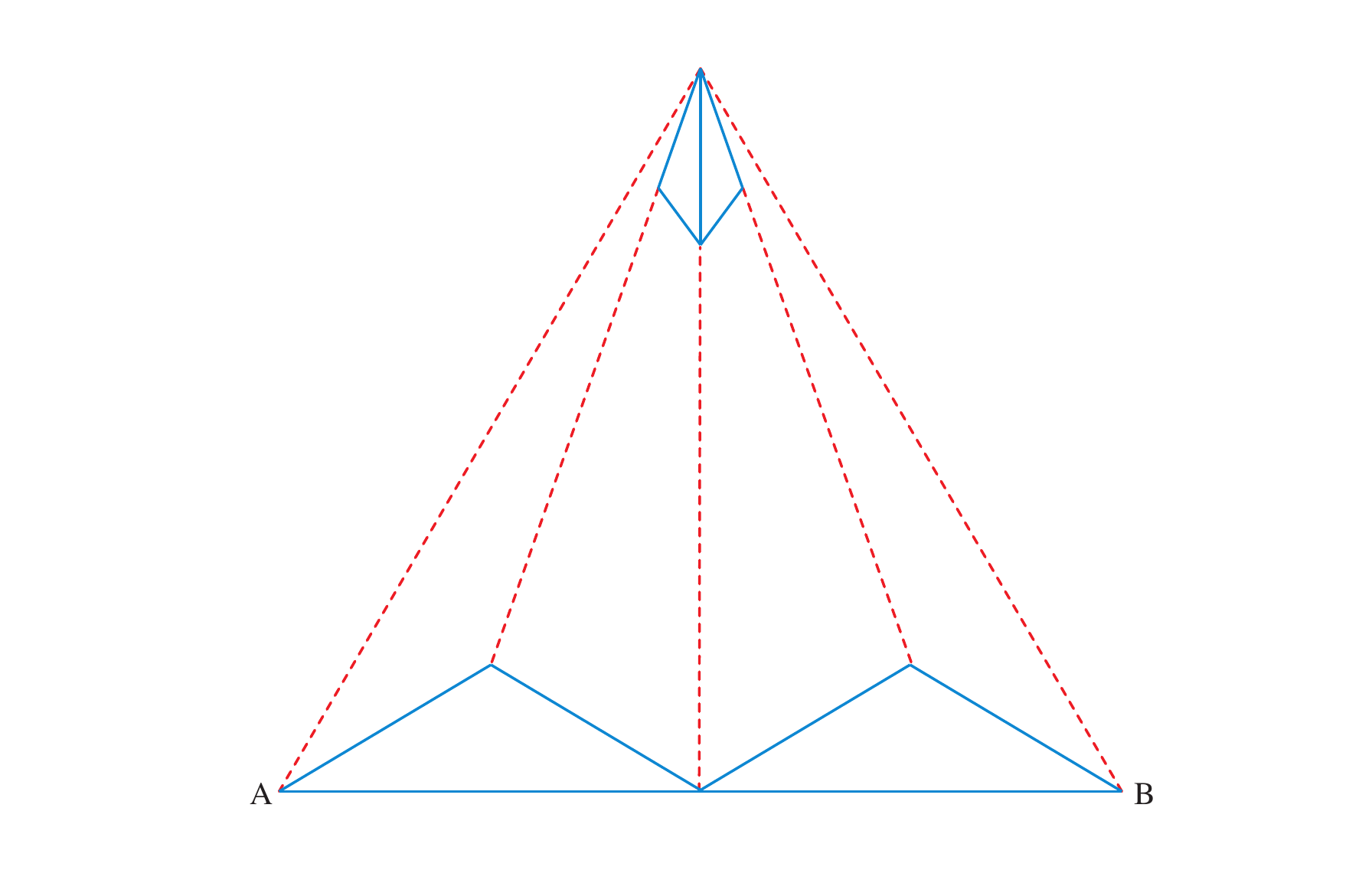}
\caption{Taking two copies of this cell complex which has the topology of a disc and identifying the boundary edges of the same color we obtain a coloured cell complex with the topology of $S^2$.  The vertices $A$ and $B$ are connected both by a red and a blue path.}
    \label{fig4}
      \end{center}
      \end{figure}

More specifically, there exist coloured  cell 
complexes as we have described that contain vertices which are 
connected by a red path, i.e., a path consisting entirely of red edges, as well as by a blue path and 
such complexes cannot be the midsection of a causal slice. 
To see this, suppose $v$ and $w$ are vertices connected by a blue path.  Then $v$ and $w$ lie on two unique  
1-simplicies with the same red endpoint in the boundary of the slice.
If they are also connected by a red path those two 1-simplices have  the same blue endpoint in the boundary.
This is only possible if $v=w$. 
It is beyond the scope of the present paper to characterize those $(D-1)$-dimensional cell 
complexes that arise as midsections of causal slices. On the other hand, it will be an important ingredient in 
the proof of 
the exponential bound to show that $\pi$ is injective.   

Before proceeding it is useful to introduce the following notation.
If $\sigma_1,\dots,\sigma_n$, are simplices whose vertices form a partition of the vertices of a simplex $\sigma$ we use the 
notation  $(\sigma_1\dots \sigma_n)$ for $\sigma$. Thus, e.g.\,, if $v$ is a vertex and $e$ is an edge then $(ve)$ denotes the triangle 
containing the edge $e$ and opposite vertex $v$. 

If $K_\phi$ is a geometric realization of an abstract simplicial complex $K$ 
in $\mathbb R^n$ 
and $\sigma$ is a simplex in $K$ we let $N_\epsilon(\sigma)$ denote the neighbourhood of $\phi( \sigma )$ 
consisting of points in $K_\phi'$ at 
distance at most $\epsilon$ from $\phi( \sigma )$. 
We shall repeatedly use the fact that if $K$ is a connected manifold of dimension $D$ 
and $S$ is a union of neighbourhoods of the form $N_\epsilon(\sigma)$ with $\sigma$ of dimension at most $D-2$ 
then $K_\phi'\setminus S$ is connected if $\epsilon>0$ is sufficiently small. Likewise, we will use the fact 
that the star of a simplex not in the boundary of $K$ 
is homeomorphic to a $D$-ball and, in particular, that any interior $(D-1)$-simplex is incident on exactly two $D$-simplices. 

\section{The 3-dimensional case}\label{3D}

In this section we first establish an exponential bound on the number of 3-dimensional causal triangulations 
introduced in Definition 1.  We then generalize this result to arbitrary boundary topology
and discuss a slightly generalized notion of causal triangulations.

\medskip
\noindent
{\bf Theorem 1.}  {\it Let $N_3(V)$ denote the number of 3-dimensional causal triangulations with $V$ tetrahedra. 
Then there exists a constant $C_3>0$ such that 
\begin{equation}\label{3bound}
N_3(V) \leq C_3^V\,,
\end{equation}
for all positive integers $V$.}

\medskip

\noindent
{\bf Proof.} The proof proceeds in three steps.

(i)~  Since the number of ways to write $V$ as a sum of positive integers is bounded by $2^V$ it suffices to find 
an exponential bound on the number of causal slices as a function of the number $V$ of tetrahedra. 

(ii)~  Consider a causal slice $K$. By assumption its interior is homeomorphic to $S^2\times (0,1)$. 
On the other hand, we have already 
remarked that it is also homeomorphic to $S_K\times (0,1)$ since the topology of the midsection does not depend on its height $h\in (0,1)$. 
In particular, $S_K\times (0,1)$ is simply connected and it follows that $S_K$ is simply connected and hence homeomorphic to $S^2$.
Now introduce a new edge for each quadrangle in $S_K$ connecting an opposite pair of its vertices and colour those edges black. This yields an 
abstract triangulation of $S^2$ with at most $2V$ triangles, since each rectangle in $S_K$ splits into two triangles. Clearly, removing the 
black edges one recovers $S_K$ and the number of possible edge colourings is bounded by $3^{3V}$, since the number of edges in a triangulation of $S^2$ with $N$ triangles equals $\frac 32 N$. 
Moreover, it is well known that the number of 
(uncoloured) abstract triangulations of $S^2$ with at most $N$ triangles is bounded by $C_0^N$, where $C_0>0$ is a constant 
(see, e.g., \cite{dur} for an elementary proof; one may also use the corresponding result for planar maps \cite{tutte}).
We conclude that the number of combinatorially distinct 
midsections $S_K$, that can occur for $|K^3|=V$, is bounded by $(27 C_0^2)^V$. 

(iii)  If $S_K$ determines $K$ up to combinatorial isomorphism, that is if the map $\pi$ 
is injective on ${\cal C}{\cal S}_3$, 
then the bound \rf{3bound}  follows from (i) and (ii). The proof of injectivity of $\pi$ is deferred to Lemma 2 
below. \hfill$\square$

\medskip
\noindent
{\bf Lemma 2.}  {\it Let $K\in {\cal C}{\cal S}_3$. Then the coloured cell complex $S_K$ determines $K$ 
up to combinatorial isomorphism.}

\medskip
\noindent
{\bf Proof.} Let $K$ be a causal slice with realization $K_\phi$ containing $S_K$.  
We will construct from $S=S_K$ a simplicial complex $K_S$, depending only on the combinatorial isomorphism class of $S_K$,  
which will be shown to be isomorphic to $K$.
  
To each vertex $v$ in $S_K$ we associate a pair of new vertices $( r_v, b_v)$ which will be in the vertex set of $K_S$.    
The vertex $ r_v$ is by assumption different from 
$b_w$ for all vertices $v,w$ in $S_K$ and $ r_v= r_w$ if and only if $v$ is connected to $w$ in $S_K$ by a blue path.
Similarly, $ b_v= b_w$ if and only if $v$ is connected to $w$ by a red path.
We note that  $ r_{v_1}=r_{v_2}$ and $ b_{v_1}= b_{v_2}$ if and only if $v_1=v_2$,
since the edges in $K_\phi$ containing $v_1$ and $v_2$ would have identical endpoints.
The vertex set $K_S^0$ of $K_S$ is by definition the set of vertices $ r_v, b_v$ where $v$ ranges over the vertices 
of $S_K$, subject to the identifications we have described. 

Now associate to each red triangle of $S_K$ with corners $v_1,v_2,v_3$  a tetrahedron of type (3,1) with red corners 
 $r_{v_1},  r_{v_2}, r_{v_3}$ and a blue corner $ b_{v_1}= b_{v_2}= b_{v_3}$.  Analogously we associate a type 
 $(1,3)$ tetrahedron to each blue triangle in $S$.   Finally, to a quadrangle cell with corners $v_1,v_2,v_3,v_4$ such 
 that $(v_1,v_2)$ and $(v_3,v_4)$ are red edges while $(v_2,v_3)$ and $(v_4,v_1)$ are blue, we associate a type 
 $(2,2)$ tetrahedron with red corners $ r_{v_1}= r_{v_4}$ and $ r_{v_2}= r_{v_3}$ and blue corners
  $ b_{v_1}= b_{v_2}$ and $ b_{v_3}= b_{v_4}$.
This defines the set of tetrahedra in $K_S$ and hence also defines $K_S$ as a 3-dimensional abstract simplicial complex 
with obvious vertex colouring. 
  
We next remark that if $v_1,v_2$ are two vertices in $S_K$ 
connected by a blue path then, as noted above, the unique edges in $K_\phi$ 
containing $v_1$ and $v_2$, respectively, have a common red endpoint which will be denoted
$\bar{r}(v_1)=\bar{r}(v_2)$.
Similarly, 
if $v_1,v_2$ are connected by a red path the two edges have a common blue endpoint 
$\bar{b}(v_1)=\bar{b}(v_2)$.
Hence, we may define a map $\tau:K_S^0\to K^0$ by setting  
$$
\tau( r_v)=\bar{r}(v) \quad\mbox{and}\quad \tau( b_v)=\bar{b}(v)\,.
$$
 It is evident by construction of $K_S$ that $\tau$ is surjective and 
maps simplices in $K_S$ to simplices in $K$. To prove that $\tau$ is a combinatorial isomorhism, 
it remains to show that $\tau$ is injective, i.e., that the defining identifications of vertices described above are the only ones.  

In order to prove this consider, say, a red vertex $r$ and the star $B_r$  in $\partial K_{red}$ 
around $r$. It is a disc whose 
boundary $\partial B_r$ is 
a circle $S^1$ consisting of red edges $e_1,\dots, e_n$, see Fig.\ 5. The boundary triangle $\Delta_i=(r e_i)$ is incident on exactly one 
\begin{figure}[h]
  \begin{center}
     \includegraphics[width=6truecm]{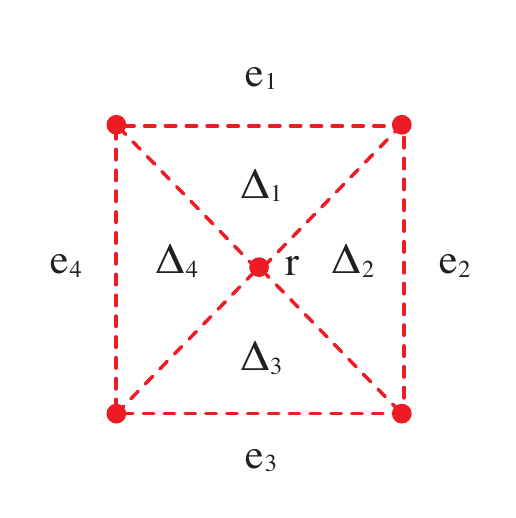}
	     \caption{The star in $\pa K_{red}$ of a red boundary vertex $r$.}
    \label{fig5}
      \end{center}
      \end{figure}
tetrahedron $t_i=(r b_i e_i)$ which is necessarly of type (3,1), as indicated, with $b_i$ denoting its blue vertex. Consider a vertex 
$v\in \partial B_r$.  We note that $r$ is not 
contained in any simplex of type (3,1) other than $t_1,\dots, t_n$, since there are no 
mono-coloured interior edges in $K$. The boundary edge $(r v)$ is contained in exactly two triangles $\Delta_i$ and $\Delta_{i+1}$ in $B_r$. 
If $b_i\neq b_{i+1}$ the triangle $(r v b_i)$ is incident on $t_i$ and one other tetrahedron $t_{i1}=(r v b_i b_{i1})$, necessarily  of type (2,2). 
If $b_{i1}\neq b_{i+1}$ the triangle $(r v b_{i1})$ is incident on $t_{i1}$ and one other tetrahedron $t_{i2}=(r v b_{i1} b_{i2})$, necessarily 
of type (2,2). Continuing, we obtain a sequence $t_{ij}=(r v b_{i(j-1)} b_{ij}), j=1,\dots,m_i$, of tetrahedra of type (2,2) such 
that $b_{i0}=b_i$ and $b_{i m_i}= b_{i+1}$. These are by construction different tetrahedra, since $b_ {i0},\dots, b_{i m_i}$ are different. 
If $b_i=b_{i+1}$ we set $m_i=0$, in which case the 
triangle $(r v b_i)$ is incident only on the tetrahedra $t_i$ and $t_{i+1}$. It is important to note that 
the edge $(r v)$ is not incident on any tetrahedra in $K$ except $t_i,t_{i 1},\dots,t_{i m_i},t_{i+1}$, since any other tetrahedron 
incident on $(r v)$  would intersect $t_i \cup t_{i 1}\cup\dots\cup t_{i m_i}\cup t_{i+1}$  only in $(r v)$ and it would follow that removing a
 neighbourhood $N_\epsilon(r v)$ from the star of $(r v)$ would yield a disconnected set for $\epsilon$ small enough, contradicting the 
manifold property of $K$. 

Repeating the above construction for all vertices $v\in\partial B_r$ we obtain a connected simplicial complex $A_0$ consisting of tetrahedra 
$t_i, t_{i1},\dots,t_{i m_i}, i=1,\dots,n$, and their subsimplices. From the preceding remark it follows that $A_0$ exhausts all the tetrahedra 
of type (3,1) or (2,2) in $K$ that are incident on $r$. We next describe how the star of $r$ in $K$ is 
obtained by successively adding tetrahedra, necessarily of type (1,3), starting from $A_0$. 

Let $C_0$ be the connected simplicial complex consisting of the blue edges and vertices of $A_0$, i.e., the 
edges of $C_0$ are $e_{ij}=(b_{i(j-1)} b_{ij}), 1\leq j\leq m_i, 1\leq i \leq n$, some of which may be identical. Pick an edge $e_{ij}$ in $C_0$. 
The triangle $\Delta_{ij}=(r e_{ij})$ is contained in $t_{ij}$ and one more tetrahedron $t$ in $K$. If $t$ is not in $A_0$ it is of type 
(1,3) and we define $A_1$ to be the simplicial complex obtained by adding $t$ (and its subsimplices) to $A_0$. Moreover, $C_1$ is defined by 
adding the triangle in $t$ opposite to $r$ to $C_0$. If $t$ is already in $A_0$ we set $A_1=A_0$ and $C_1=C_0$. Next repeat the same construction 
for any edge $e$ in $C_1$ to obtain $A_2$ and $C_2$ and continue until the complex $A_N$ and its blue subcomplex $C_N$ are obtained with the 
property that any triangle $(r e)$, where $e$ is an edge in $C_N$, is incident on exactly two tetrahedra in $A_N$.      

Suppose there exists a tetrahedron $t'$ of type (1,3) incident on $r$ which is not in $A_N$. Then $t'$ is not incident on any triangle in $A_N$ 
by construction and hence it can share at most two edges with $A_N$. It follows that by removing a sufficiently small neighbourhood of 
the 1-skeleton of the star of $r$ in $K$ one would obtain a disconnected set, contradicting the manifold property of $K$. Thus we have shown that
$A_N$ equals the star $\widetilde B_r$ of $r$ in $K$. Moreover, the blue subcomplex $C_N$ of $A_N$ is by construction connected and its vertices are
 exactly the set of blue vertices that occur as endpoints of edges in $K$ whose other endpoint is $r$. It follows that any two midpoints of 
those edges in $S_K$ are connected by a blue path. 
Applying an analogous argument for the blue vertices $b\in\partial K_{blue}$, we conclude that $\tau$ 
is injective and this completes the proof that $\tau$ is a combinatorial isomorphism.

It is clear from the construction of $K_S$ that it depends only on the combinatorial isomorphism class of $S=S_K$. It therefore 
follows from the existence of $\tau$ that $S_K$ determines $K$ up to combinatorial isomorphism as claimed. 
\hfill $\square$

\bigskip

\noindent
{\bf Remark.} An alternative route to Theorem 1 would be to show that every causal slice has a local construction 
in the sense of \cite{dj}.  While we believe that this is feasible, we find the argument using the
midsection $S_K$ more transparent.

\medskip

The exponential bound in Theorem 1 and some standard subadditivity arguments allow us to get a sharper estimate 
on the growth of the number of $3$-dimensional causal triangulations with fixed boundaries as we now explain.
Let ${\cal T}(\Sigma_{\rm in},\Sigma_{\rm out})$ be the set of all (combinatorial isomorphism classes of) 3-dimensional causal triangulations with boundary components $\Sigma_{\rm in}$ and $\Sigma_{\rm out}$.  Let $N(V,\Sigma_{\rm in},\Sigma_{\rm out})$ denote the number of triangulations in ${\cal T}(\Sigma_{\rm in},\Sigma_{\rm out})$ of volume $V$, i.e., 
consisting of $V$ tetrahedra.   

\medskip

\noindent
{\bf Theorem 2.}  {\it The limit
$$
\lim_{V\to\infty}{\log N(V,\Sigma_{\rm in},\Sigma_{\rm out})\over V}
$$
exists for all $\Sigma_{\rm in}$, $\Sigma_{\rm out}$ and is independent of $\Sigma_{\rm in}$ and $\Sigma_{\rm out}$.}

\medskip

In order to prove this Theorem we need the following simple Lemma which we assume for a moment.

\medskip

\noindent
{\bf Lemma 3.} {\it The sets ${\cal T}(\Sigma_{\rm in},\Sigma_{\rm out})$ are all nonempty.}

\medskip

\noindent
{\bf Proof of Theorem 2.}   Let 
$\Sigma_{\rm in}$ and $\Sigma_{\rm out}$ be two triangulations of $S^2$.  Choose a causal triangulation $T_0\in
 {\cal T}(\Sigma_{\rm in},\Sigma_{\rm out})$ and let $V_0$ be the volume of $T_0$.  
 Let $T_1$ and $T_2$ be two causal triangulations in the same set with 
 volumes $V_1$ and $V_2$.  
   Now glue $T_1$ to $T_0$ along $\Sigma_{\rm out}$ and glue $T_2$ to $T_0$ along $\Sigma_{\rm in}$, using arbitrarily chosen combinatorial isomorphisms to identify boundary components.  
   Then we obtain a new element $T'$ in ${\cal T}(\Sigma_{\rm in},\Sigma_{\rm out})$ of volume $V_1+V_2+V_0$.  
   Clearly $T_1$ and $T_2$ are uniquely determined by $T'$ and it follows that
 $$
 N(V_1,\Sigma_{\rm in},\Sigma_{\rm out})N(V_2,\Sigma_{\rm in},\Sigma_{\rm out})\leq N(V_1+V_2+V_0, \Sigma_{\rm in}, \Sigma_{\rm out}).
 $$
 Hence, the function
 $$
 f(V)=-\log N(V-V_0,\Sigma_{\rm in},\Sigma_{\rm out})
 $$
 is subadditive (for $V$ sufficiently large) so (see, e.g., \cite{hp} Sec.\ 7.6)
 $$
- \lim_{V\to\infty}{f(V)\over V}=-\inf_{V}{f(V)\over V}\equiv \beta (\Sigma_{\rm in},\Sigma_{\rm out})
 $$
 which is finite by Theorem 1.  A priori $\beta (\Sigma_{\rm in},\Sigma_{\rm out})$ depends on the boundary triangulations $\Sigma_{\rm in}$ and $\Sigma_{\rm out}$.  However, given an arbitrary causal triangulation in $ {\cal T}(\Sigma_{\rm in},\Sigma_{\rm out})$ 
 we can glue on it two fixed triangulations $T_1$ and $T_2$ from $ {\cal T}(\Sigma_{\rm in}',\Sigma_{\rm in})$ and 
 $ {\cal T}(\Sigma_{\rm out},\Sigma_{\rm out}')$ for any triangulations $\Sigma_{\rm in}'$ and $\Sigma_{\rm out}'$ of the 2-sphere.  It follows that
 $$
 N(V,\Sigma_{\rm in},\Sigma_{\rm out})\leq N(V+V_1+V_2, \Sigma_{\rm in}',\Sigma_{\rm out}'),
 $$
 where $V_i$ is the volume of $T_i$, $i=1,2$.   We deduce that
 $$
 \beta (\Sigma_{\rm in},\Sigma_{\rm out})\leq \beta (\Sigma_{\rm in}',\Sigma_{\rm out}')
 $$
 for any two pairs of triangulations of $S^2$.  Hence, all the $\beta$'s are equal and we have established 
 the Theorem.   \hfill  $\square$

\medskip
\noindent
{\bf Proof of Lemma 3.}  Let $\Sigma_T$ be the boundary of a tetrahedron.  
We will show that for any triangulation $\Sigma$ of the 2-sphere there is a causal slice with boundary 
components $\Sigma$ and
$\Sigma_T$.  Gluing two such causal slices along the boundary tetrahedron we can then obtain a causal 
triangulation with any prescribed in- and out-boundaries.

 Consider the cone $C_\Sigma$ over $\Sigma$, that is to say the 
3-dimensional simplicial complex 
constructed by introducing one new vertex $a$ and declaring the tetrahedra to be $(\Delta a)$ where $\Delta$ ranges over the triangles in $\Sigma$.  The tetrahedra $(\Delta a)$ and $(\Delta 'a)$ are glued together along the triangle $(\ell a)$ if and only if the two triangles $\Delta$ and $\Delta '$ in $\Sigma$ are glued along the 
edge $\ell$. 

Now suppose for simplicity that $\Sigma$ has a vertex $0$ of degree three.  
Denote the neighbours of $0$ by 1,2 and 3.  Consider the star of the vertex $0$ in the cone $C_\Sigma$ which consists of the tetrahedra $(012a)$, $(013a)$ and $(023a)$.   Remove these three tetrahedra from $C_\Sigma$.  Take a copy of $\Sigma_T$ and identify one of the vertices with $a$.  Label the other vertices $b,c,d$.   Colour the vertices and edges in 
$\Sigma$ red and those in $\Sigma_T$ blue.  Now we fill up this simplicial complex so that we get a causal slice.  First introduce the $(3,1)$ tetrahedra $(b023)$, $(c013)$ and $(d012)$ which fill the hole in $\Sigma$.  Then we glue $(1,3)$ tetrahedra to the $\Sigma_T$.  These are: $(acd1)$, $(abd2)$, $(abc3)$ and $(bcd0)$.  It remains to glue the 
$(1,3)$ tetrahedra to the $(3,1)$ tetrahedra.  For this purpose we need 6 $(2,2)$ tetrahedra which are $(cd01)$, $(bd02)$, $(bc03)$, 
$(ac13)$, $(ad12)$ and $(ab23)$, see Fig.\ 6 which illustrates part of the midsection of the causal slice we have 
constructed.  
\begin{figure}[h,t]
  \begin{center}
    \includegraphics[width=9truecm]{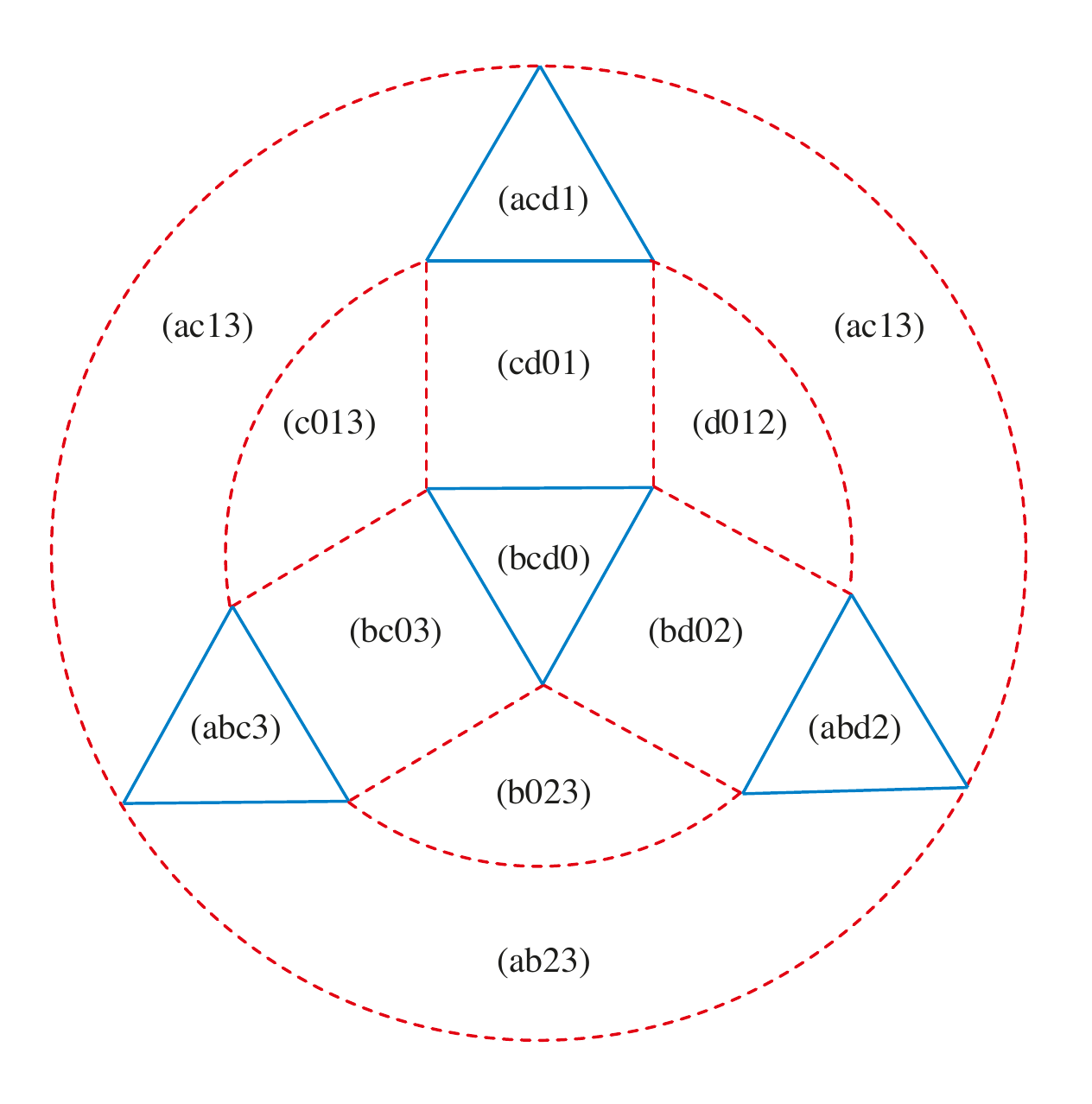}
 \caption{The part of the midsection describing the gluing of the 13 tetrahedra replacing the star of $0$ in 
$C_\Sigma$.   Here $(ac13)$ refers to the tetrahedron with vertices $a,c,1,3$ etc.}
    \label{fig9}
      \end{center}
      \end{figure}
      
In this construction we began by removing 3 tetrahedra from $C_\Sigma$ and then we added 13 tetrahedra to get a causal slice with boundary components $\Sigma_T$ and $\Sigma$.  Hence, for any two triangulations $\Sigma_{\rm in}$ and $\Sigma_{\rm out}$ of $S^2$ 
there is a causal 
triangulation of volume $|\Sigma_{\rm in}|+|\Sigma_{\rm out}|+20$ with boundary components $\Sigma_{\rm in}$ and $\Sigma_{\rm out}$.   Here $|\Sigma_{\rm in}|$ is the number of triangles in $\Sigma_{\rm in}$ and similarly for  $\Sigma_{\rm out}$.

If $\Sigma$ has no vertex of degree 3, it necessarily has a vertex of order 4 or 5.  It is easy modify the construction above to this case by subdividing some of the tetrahedra.  We leave the details to the reader.
\hfill $\square$

\medskip

In the remainder of this section we discuss 3-dimensional causal triangulations with boundary components 
of arbitrary topology.  We also relax the requirement of cylindrical topology in condition (i) of Definition 1. 

\medskip
\noindent
{\bf Definition 2.}  A $D$-dimensional {\it generalized causal slice} $K$ is an abstract 
simplicial complex which satisfies the following 
conditions:
\begin{enumerate}

\item[(i)] $K$ is an oriented $D$-manifold with two boundary components.  

\item[(ii)] All mono-coloured simplices of $K$ belong to the boundary, such that the red ones belong to one 
boundary component $\partial K_{red}$ and the blue ones to the other component $\partial K_{blue}$.  
 \end{enumerate}
A {\it generalized causal triangulation} of dimension $D$ is an abstract simplicial complex of the form 
$$
M = \bigcup_{i=1}^N K^i\,,
$$
where each $K^i$ is a $D$-dimensional generalized causal slice such that $K^i$ and $K^j$ are disjoint if $i\neq j$ except 
that $\partial K^{i}_{blue} = \partial K^{i+1}_{red},\, i= 1,\dots, N-1,$ as 
uncoloured abstract simplicial complexes. The boundary components of $\partial K^1_{red}$ and $\partial K^N_{blue}$ of $K$ 
will be denoted $\Sigma_{in}$ and $\Sigma_{out}$, respectively.  

\medskip

\noindent
{\bf Lemma 4.} {\it For any generalized $3$-dimensional causal slice $K$ 
the midsection $S_K$ is homeomorphic to both $\partial K_{red}$ and 
$\partial K_{blue}$.  }

\medskip
\noindent
{\bf Remark.}  Before proving the above lemma we note that if we assume that $K$ is a 
topological cylinder, i.e., homeomorphic to $[0,1]\times \Sigma$ for some two-dimensional orientable manifold 
$\Sigma$, then the statement of Lemma 4 is trivial since we know that the interior of $K$ is
homeomorphic to $(0,1)\times S_K$ and hence the homotopy group $\pi_0(S_K)$ is isomorphic to $\pi_0(\Sigma)$  and 
this implies that $S_K$ is homeomorphic to $\Sigma$.  

\medskip

\noindent
{\bf Proof.} Let $K$  be a generalized causal slice. We use the same notation as in the proof of Lemma 2 and consider 
the star $\widetilde B_r$ in $K$ around a red vertex 
$r\in \partial K_{red}$ consisting of type (3,1) tetrahedra 
$t_i$ and type (2,2) tetrahedra $t_{ij}$ and possibly 
additional tetrahedra of type (1,3).  

Since $K$ is an oriented manifold it is clear that $S_K$  is an orientable surface. It therefore suffices to show that the 
Euler characteristics of $\partial K_{red}$ and $S_K$ coincide, the argument for $\partial K_{blue}$ being similar.

Consider the intersection of $\widetilde B_r$ with $S_K$.  Its boundary $\partial (\widetilde B_r\cap S_K)$ in 
$S_K$ is a simple curve $\mu_r$ consisting 
of edges each of which belongs to either a quadrangle $t_{ij}\cap S_K$ and is blue or to a red 
triangle $t_i\cap S_K$. Since 
$\partial\widetilde B_r\setminus \partial K_{red}$ is a disc, it follows 
by the manifold property of $K$ that $\mu_r$ bounds a disc in  
$\partial\widetilde B_r\setminus \partial K_{red}$ and this implies that $\widetilde B_r\cap S_K$ is a disc containing the quadrangles 
$t_{ij}\cap S_K$ and the triangles $t_i\cap S_K$. 

Next, we define a graph $G_{red}$ in $S_K$ whose vertices are the barycenters of quadrangles or red triangles in $S_K$. The edges of 
$G_{red}$ are dual to red edges in $S_K$ in the following sense: each red edge $e$ in $S_K$ is incident on two cells, whose 
barycenters $\alpha$ and $\beta$ are vertices of $G_{red}$. 
Let $a_e$ be the arc in $S_K$ consisting of the two line segments connecting 
the barycenter of $e$ to $\alpha$ and $\beta$, respectively, see Fig.\ 7.

\begin{figure}[h,t]
  \begin{center}
      	\includegraphics[width=4truecm]{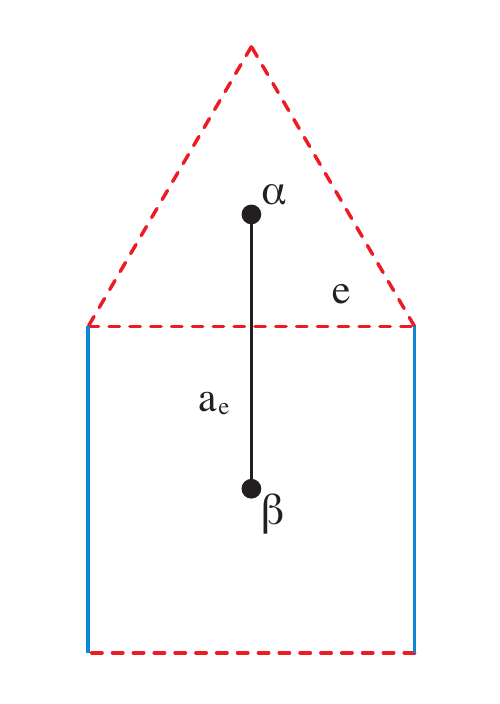}
  \caption{An edge in $G_{red}$ joining the centers of a red triangle and a quadrangle.}
    \label{fig6}
      \end{center}
      \end{figure}

 Then $a_e$ connects $\alpha$ and $\beta$ and, clearly, the interiors of $a_e$ 
and $a_f$ are disjoint for any two different red edges $e$ and $f$. Hence $G_{red}$ is defined as a graph in $S_K$ by letting its edge set 
consist of the arcs $a_e$, where $e$ is any red edge in $S_K$. Note that a vertex of $G_{red}$ is of order $2$ or $3$ depending on whether 
it is the barycenter of a quadrangle or a red triangle, respectively, see Fig.\ 8.
\begin{figure}[h]
  \begin{center}
      	\includegraphics[width=8truecm]{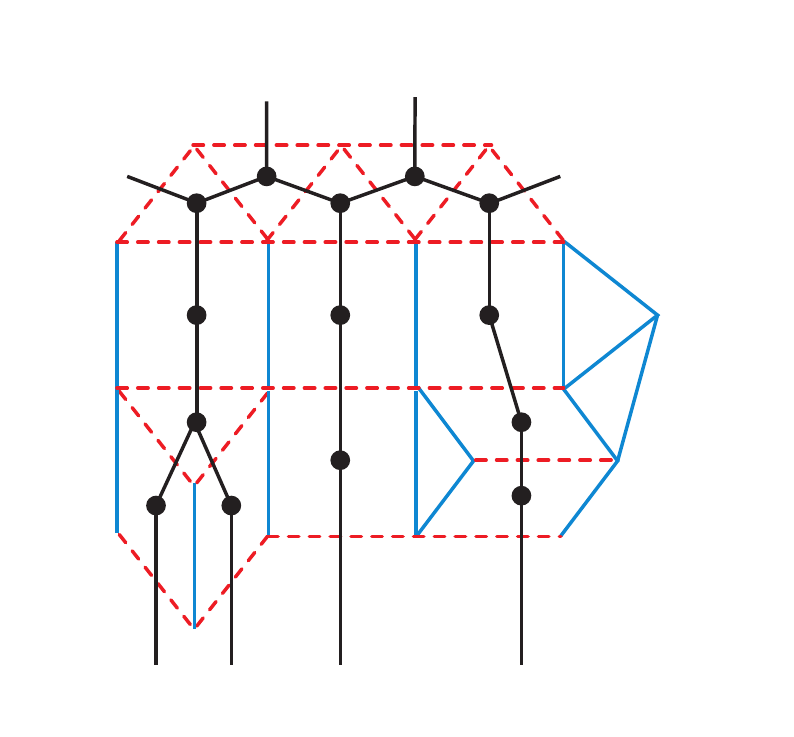}
 \caption{The graph $G_{red}$ corresponding to the midsection in Fig.\ 2.}
    \label{fig7}
      \end{center}
      \end{figure}
      Since the red boundary component 
$\pa K_{red}$ is connected it follows that the graph $G_{red}$ is connected. 

Returning to the disc $\widetilde B_r\cap S_K$, we see that it contains the barycenters of the cells $t_{ij}\cap S_K$ and $t_i\cap S_K$ 
as well as the circuit $\gamma_r$ in $G_{red}$ made up of the edges connecting them. Clearly, $\gamma_r$ is a deformation of $\mu_r$ 
and likewise bounds an open disc $D_r$ in $S_K$, see Fig.\ 9.

\begin{figure}[h]
  \begin{center}
      	 \includegraphics[width=10truecm]{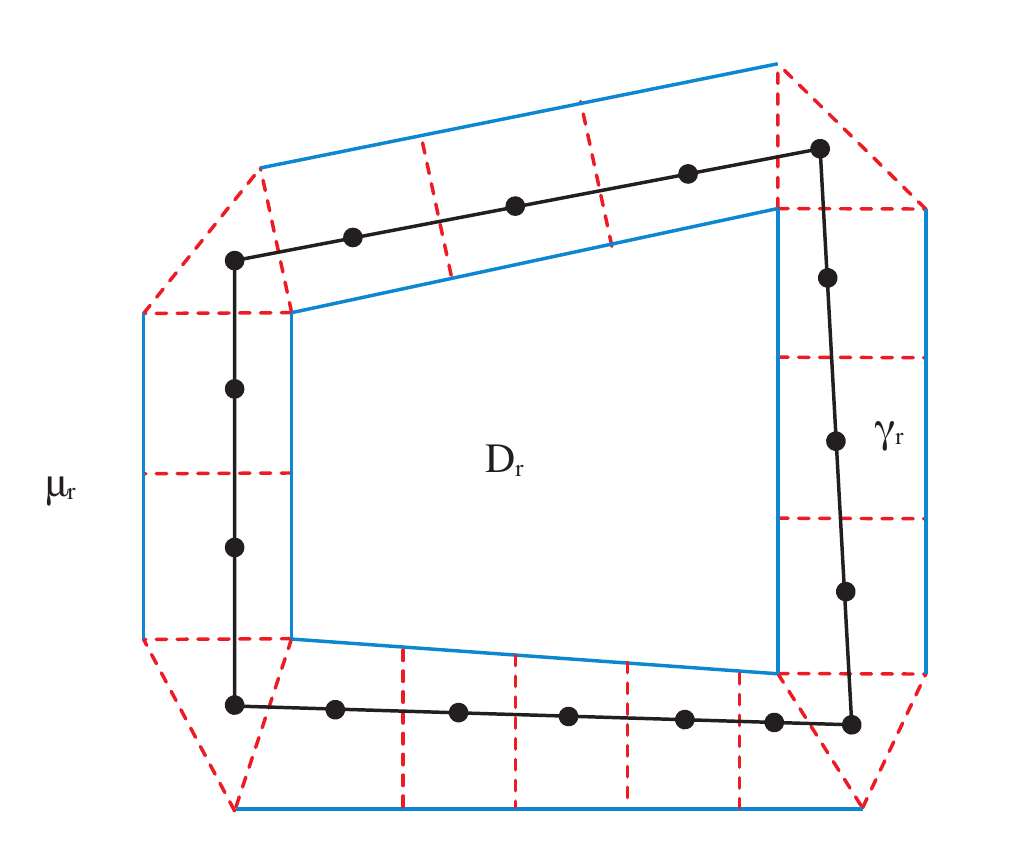}
\caption{The curves $\mu_r$ and $\gamma _r$ and the disc $D_r$.  In this case the red vertex $r$ has order 4.}
    \label{fig8}
      \end{center}
      \end{figure}

By the construction of $\widetilde B_r$ in the proof of Lemma 2, the disc 
$D_r$ contains no part of $G_{red}$, and 
$D_r\cap D_{r'}=\emptyset$ for $r\neq r'$. Moreover, given a red edge $e$ in $S_K$ it is contained in a unique triangle in $K$ 
that contains a single edge $e'$ in $\partial K_{red}$. Letting $r$ and $r'$ be the endpoints of $e'$ it follows that $a_e$ 
belongs to $\gamma_r$ and $\gamma_{r'}$. This shows that the complement of $G_{red}$ in $S_K$ is a disjoint union of discs labelled by 
the vertices of $\partial K_{red}$. Hence, the Euler characteristic $\chi(S_K)$ 
of $S_K$ can be calculated by Euler's formula in terms of the graph $G_{red}$.
If $F$ is the number of faces of $G_{red}$, $E$ the number of edges and $V$ the number of vertices we have
$$
\chi(S_K) = V-E+F\,.
$$
The number of faces of $G_{red}$ equals the number of vertices in $\partial K_{red}$ by the previous remark and 
the  number of edges in 
$G_{red}$ equals the number of red edges in $S_K$ and fulfills
\begin{equation}\label{12345}
2E=3\#\{ \mbox{\rm red triangles in } S_K\} + 2\#\{ \mbox{\rm quadrangles in } S_K\}\,.
\end{equation}
Finally, the number of vertices in $G_{red}$ equals
\begin{equation}\label{vert}
V= \#\{ \mbox{\rm red triangles in} S_K\} + \#\{ \mbox{\rm quadrangles in } S_K\}\,.
\end{equation}
It follows that 
\bea
\chi (S_K) & = & \#\{ \mbox{\rm vertices in } \partial K_{red}\} -\oh \#\{ \mbox{\rm red triangles in } S_K \}\nonumber\\
         & = & \chi (\partial K_{red})
\eea
since $\#\{ \mbox{\rm red triangles in } S_K\}=\#\{ \mbox{\rm triangles in } \partial K_{red}\}$ and 
\begin{equation}
2\#\{ \mbox{\rm edges in } \partial K_{red}\} =3 \#\{ \mbox{\rm triangles in } \partial K_{red}\} .
\end{equation}
This completes the proof.   \hfill $\square$

\medskip
\noindent
{\bf Remark.}  A proof of the lemma above can also be based on the existence of collar neighbourhoods of the boundary components (see, e.g., Corollary 2.26 in \cite{plt}), but we prefer to give the preceding more direct proof. 

It can presumably be proven that a $3$-dimensional generalized causal slice $K$ is homeomorphic to a cylinder. But we shall not need this fact and will not elaborate further on it here, except for noting that in case the boundary components $\partial K_{\rm red}$ and $\partial K_{\rm blue}$ are homeomorphic to $S^2$ this fact follows easily from the validity of the Poincar{\'e} conjecture \cite{perel}. Indeed, 
 it follows from Lemma 4 that the interior of $K$ is homeomorphic to $(0,1)\times S^2$. 
In particular, the interior of $K$ is simply connected. Being a manifold, $K$ has the structure of a cylinder close to its boundary by the collaring theorem. It follows that $K$ is also simply 
connected. By gluing two triangulated 3-balls onto the two boundary components of $K$ one therefore obtains a simply connected closed 
3-manifold which is homeomorphic to the 3-sphere by the Poincar{\'e} conjecture. Removing the interior of the two 3-balls shows that $K$ is homeomorphic to a cylinder $[0,1]\times S^2$. Hence, in dimension 3 the notion of a generalized causal 
slice with boundary components homeomorphic to $S^2$ is equivalent to that of a causal slice. 

\medskip

Henceforth, we denote by ${\cal C}_{3,g}$ the set of generalized causal 
triangulations whose boundary components have genus $g$ and we let
$N_{3,g}(V)$ denote the number of $K$ in ${\cal C}_{3,g}$ such that $|K^3|=V$. We then have the following generalization of Theorem 1.

\medskip
\noindent
{\bf Theorem 3.} {\it There exists a constant $C_{3,g}>0$ such that 
$$
N_{3,g}(V) \leq C_{3,g}^V\,,
$$
for all positive integers $V$.
}

\noindent
{\bf Proof.} It is sufficient to verify the three steps in the proof of Theorem 1. Step (i) requires no change and the same 
applies to step (iii) since the topology of $S_K$ was not used to prove injectivity of $\pi$. In step (ii) the only modification 
required is the establishment of a bound on the number of triangulations of an oriented surface of fixed genus $g$ as a function 
of the number of triangles. This result may be found in \cite{dur} or, alternatively, it can easily be deduced 
from \cite{tutte}.\hfill $\square$

\medskip
\noindent
{\bf Remark.}  We note that the arguments in the proof of Theorem 2 apply to the case of boundaries with
nonzero genus except that a suitable generalization of Lemma 3 is missing.

\section{The 4-dimensional case}\label{4D}

In this section we  discuss the extension of Theorem 1 to the case of 4-dimensional 
causal triangulations. We let $M_3(V)$ denote the number of all abstract simplicial complexes homeomorphic to the 3-sphere made up of $V$ 
tetrahedra and denote by $N_4(V)$ 
the number of 4-dimensional causal triangulations (as defined in Definition 1) made up of $V$ 4-simplices. 
 Our main result is the following.

\medskip
\noindent
{\bf Theorem 4.} {\it If there exists a constant $C>0$ such that $M_3(V)\leq C^V$ for all $V\geq 0$, then 
there exists a constant $C_4>0$ such that
$$
N_4(V) \leq C_4^V\,,
$$
for all positive integers $V$.
}

\noindent
{\bf Proof.} We proceed in four steps. 

(i)~  By the same argument as in the proof of Theorem 1 it suffices to bound the number $N_4'(V)$ of causal slices with $V$ 4-simplices.

(ii)~  Consider a causal slice $K$. By Lemma 1, the midsection $S_K$ is a closed 3-manifold and the interior of $K$ is homeomorphic 
to $S_K\times (0,1)$ as well as to $S^3\times (0,1)$. 
In particular, $S_K\times (0,1)$ is simply connected and it follows that $S_K$ is simply connected and hence homeomorphic to $S^3$ by 
the Poincar{\'e} conjecture \cite{perel}. 

(iii)~ As explained in Section~\ref{prelim} the coloured cell complex $S_K$ has 3-cells of four types: red or blue tetrahedra and prisms 
whose end triangles are either both red or both blue. It is well known (see e.g. \cite{plt} Proposition 2.9) that any cell complex can be 
subdivided to a simplicial complex without introducing new vertices. Let $S_K'$ be such a subdivision of $S_K$ and colour its new edges black. 
We note that each such edge $e$ connects two opposite vertices in a rectangular face $f_e$ of a  prism of $S_K$ and that each prism $p$ of $S_K$ is 
subdivided in three tetrahedra all of which are incident on one of the three black edges in the boundary of $p$. 
In order to see that $S_K$ can be reconstructed from the coloured simplicial complex $S_K'$ we further note that any black edge $e$ of $S_K'$ 
is incident on exactly two triangles $\Delta_e, \Delta_e'$ whose other edges are not black, and these triangles make up the face $f_e$.  If 
 $e$ is incident on three tetrahedra on one side of $f_e$ then those tetrahedra form a subdivision of a prism with  $f_e$ as a face. 
These observations serve to show that the colouring of $S_K'$ allows a reconstruction of the prisms of $S_K$ which together with the monocoloured 
tetrahedra define the cell complex $S_K$. 

 Noting that the number of 3-cells in $S_K$ equals $V$ and that the number of tetrahedra of $S_K'$ is bounded 
 by $3V$,  an exponential bound on 
the number of coloured cell complexes $S_K$ as a function of $V$ will follow from an 
exponential bound on the number of coloured simplicial 
complexes $S_K'$ as a function of the number of $tetrahedra$.   We obtain such a bound from (ii) 
and the assumed exponential bound on $M_3(V)$ since 
the number of possible colourings of the edges of a given simplicial manifold made up of $V$ tetrahedra is bounded by $9^V$.

(iv)  If $S_K$ determines $K$ up to combinatorial isomorphism, that is if the map $\pi$ is injective on ${\cal C}{\cal S}_4$, 
the claimed bound now follows from (i) and (iii). The proof of injectivity of $\pi$ is deferred to Lemma 5 below.      \hfill $\square$

\medskip
\noindent
{\bf Lemma 5.}  {\it Let $K\in {\cal C}{\cal S}_4$. Then the coloured cell complex $S_K$ determines $K$ up to combinatorial isomorphism.}

\medskip
\noindent
{\bf Proof.} Given a causal slice $K\in{\cal C}{\cal S}_4$ with realization $K_\phi$ we construct from $S=S_K$ a 4-dimensional simplicial 
complex $K_S$ in complete analogy with the construction in the proof of Lemma 2 and whose details are left to the reader. The proof of the Lemma
is complete once we show that $K_S$ is combinatorially isomorphic to $K$ and this in turn requires showing that if if $v_1, v_2$ are vertices 
of $S_K$ and $\bar{r}(v_1)=\bar{r}(v_2)$, using notation as in the proof of Lemma 2, then $v_1$ and $v_2$ are connected by a blue path and, 
similarly, if $\bar{b}(v_1)=\bar{b}(v_2)$ then $v_1$ and $v_2$ are connected by a red path. The proof of this fact follows by the same line of 
argument as in the the 3-dimensional case as we now describe.

Let $r$ be a vertex in $\partial K_{red}$ and consider the star $B_r$ of $r$ in $\partial K_{red}$, which is a 3-ball whose boundary $\partial B_r$ 
is a triangulated 2-sphere the triangles of which we denote by $\Delta_1,\dots,\Delta_n$. The boundary tetrahedron $t_i=(r\Delta_i)$ is 
incident on a unique 4-simplex $s_i=(r b_i\Delta_i)$, and $r$ is not contained in any other 4-simplex of type (4,1) except $s_1,\dots,s_n$, since no mono-coloured interior edges exist in $K$. 

Consider an edge $e$ in $\partial B_r$. The boundary triangle $(re)$ is contained in exactly two tetrahedra $t_{e1}=t_i$ and $t_{e2}=t_j$ in $B_r$. 
If $b_i\neq b_j$ we repeat the construction applied in the 3-dimensional case to obtain a sequence $s_{e1},\dots, s_{em_e}$ of 4-simplices of 
type (3,2) containing $(re)$, such that $s_{ei}=(r b_{e(j-1)} b_{ej} e), j=1,\dots, m_e$, and $b_{e0}=b_i$ and $b_{em_e}=b_j$. These are by 
construction different tetrahedra. Moreover, the triangle $(re)$ is not contained in any other 4-simplex of type (3,2) in $K$. In order to 
see this, observe that any 
such 4-simplex would intersect $s_i\cup s_{e1}\cup\dots\cup s_{em_e}\cup s_j$ only in $(re)$ by the assumption that any tetrahedron $(r b_{ek} e)$ 
is contained in exactly two 4-simplices. It would follow that removing a small neighbourhood $N_\epsilon(re)$ from the star of $(re)$ in $K$ 
one would obtain a disconnected set, contradicting the manifold property of $K$. 

 Repeating this construction for all edges $e$ in $\partial B_r$ we let $A_0$ denote the resulting simplicial complex consisting of the 
4-simplices $s_i, i=1,\dots, n$, and $s_{ek}, k=1,\dots,m_e$, and their subsimplices.  We let $C_0$ be the simplicial complex 
consisting of the edges $(b_{e(k-1} b_{ek})$ and their vertices. We note that $C_0$ is a connected simplicial complex that equals the blue 
subcomplex of $A_0$.
If we remove a sufficiently small neighbourhood of the 2-skeleton of $A_0$ then a connected set remains. These properties are 
preserved under the subsequent constructions.

Pick now a vertex $v$ in $\partial B_r$ and let $f$ be an edge in $C_0$. If the 
tetrahedron $t=(r v f)$ is contained in exactly one 4-simplex $s$ in $A_0$, then we define $A_1$ by adding 
to $A_0$ the other 4-simplex $s'$ in $K$ that is 
incident on $t$.  We let $C_1$ be obtained from $C_0$ by adding the triangle in $s'$ opposite to $(rv)$. Clearly, 
this triangle contains $f$. If, on the other hand, either $t$ is not in $A_0$ or $t$ is incident on two 4-simplices in $A_0$ we set $A_1=A_0$ 
and $C_1=C_0$. Now repeat this step for some edge $f$ in $C_1$ to obtain $A_2$ and $C_2$ and continue until
 all tetrahedra in $A_N$ 
of the form $(rvf)$, where $f$ is an edge in $C_N$, are incident on two 4-simplices in $A_N$. Then no further 4-simplices of type (2,3) 
in $K$ are incident on $(rv)$. In fact, such a simplex $(rv\Delta)$ would at most share three vertices with $C_N$ and hence at most three triangles 
with $A_N$, and we can use the same argument as above to conclude that the star of $(rv)$ in $K$,  with a small neighbourhood of its 
2-skeleton removed, would be disconnected, contradicting the manifold property of $K$.

Now repeat the construction for all vertices $v$ in $\partial B_r$ to obtain $A_M$ and $C_M$ such that $A_M$ contains all 4-simplices of 
types (4,1), (3,2) and (2,3) in the star of $r$ in $K$  and such that $C_M$ is connected and equals the blue subcomplex of $A_M$.
Finally, consider a triangle $\Delta$ in $C_M$. Then the tetrahedron $(r\Delta)$ is contained in some 4-simplex of $A_M$ by construction. If 
the other 4-simplex of $K$ incident on $(r\Delta)$ is not in $A_M$ define $A_{M+1}$ by adding it to $A_M$ and define $C_{M+1}$ by adding its 
blue tetrahedron to $C_M$. Otherwise, $A_M$ and $C_M$ are left unchanged. Now continue repeating this construction untill $A_L$ and $C_L$ are 
obtained such that both 4-simplices in $K$ incident on $(r\Delta)$ are contained in $A_L$ for all triangles $\Delta$ in $C_L$, the blue 
subcomplex of $A_L$. 

Then $A_L$ is the star of $r$ in $K$: otherwise, there would be further 4-simplices of type (1,4) in $K$ containing $r$ and these could at most 
share a union of triangles $(re)$ or edges $(rv)$, where $e$ is an edge in $C_L$ and $v$ is a vertex in $C_L$, with $A_L$,  and a contradiction is 
obtained in the same ways as previously.

Since $C_N$ is connected and its vertices coincide with the endpoints of edges originating from $r$ it follows that $\bar r(v_1)=\bar r(v_2)=r$ 
implies that $v_1$ and $v_2$ are connected by a blue path. Clearly, an analogous argument applies to any blue vertex $b$ instead of $r$ and 
the proof in complete.    \hfill $\square$

\medskip

\noindent
{\bf Acknowledgements.}  This work was partly supported by the NordForsk researcher network "Random Geometry" (33000).  We would like to thank S{\o}ren Galatius, Mauricio E. G. Lopez and Erik Kj{\ae}r Pedersen for helpful discussions on piecewise linear topology.  We are grateful to Ingi Freyr Atlason for help with the pictures.

\end{document}